\documentclass[12pt]{iopart}
\usepackage[UKenglish]{babel}
\usepackage{graphics}
\usepackage{tikz}
\usepackage{caption}
\usepackage{subcaption}
\usepackage{cases}
\usepackage{iopams}
\usepackage{url}
\usetikzlibrary{fadings}

\usepackage{amsfonts}
\usepackage{cases}
\usepackage{xspace}
\usepackage{psfrag}
\usepackage{subfig}
\usepackage{color}
\newcommand{\ave}[1]{\left\langle #1 \right\rangle}

\newcommand{\dint}[1]{\mathchoice{\!\mathrm{d}#1\,}{\!\mathrm{d}#1\,}{\!\mathrm{d}#1\,}{\!\mathrm{d}#1\,}}

\usepackage{dsfont}

\newcommand{\gpset}[1]{\mathds{#1}}

\newcommand{\canetset}[1]{{\mathchoice {\hbox{$\sf\textstyle #1\kern-0.4em #1$}}
{\hbox{$\sf\textstyle #1\kern-0.4em #1$}}
{\hbox{$\sf\scriptstyle #1\kern-0.3em #1$}}
{\hbox{$\sf\scriptscriptstyle #1\kern-0.2em #1$}}}}

\newcommand{\Nset}{\gpset{N}}

\def\nbZ{{\mathchoice {\hbox{$\sf\textstyle Z\kern-0.4em Z$}}
{\hbox{$\sf\textstyle Z\kern-0.4em Z$}}
{\hbox{$\sf\scriptstyle Z\kern-0.3em Z$}}
{\hbox{$\sf\scriptscriptstyle Z\kern-0.2em Z$}}}}
\newcommand{\gpvec}[1]{\mathbf{#1}}

\newcommand{\xvec}{\gpvec{x}}

\newcommand{\erf}{\operatorname{erf}}

\newcommand{\half}{\mathchoice{\frac{1}{2}}{(1/2)}{\frac{1}{2}}{(1/2)}}

\newcommand{\elabel}[1]{\label{#1}}

\newcommand{\flabel}[1]{\label{#1}}
\newcommand{\latin}[1]{{\it #1}}
\newcommand{\ie}{\latin{i.e.}\@\xspace}
\newcommand{\eg}{\latin{e.g.}\@\xspace}

\newcommand{\keyword}[1]{{\bf #1}}
\newlength \standardfigwidth
\DeclareMathAlphabet{\matheub}{U}{eur}{m}{n}

\newcounter{exercise}
{\addtocounter{exercise}{1}\begin{center}\begin{minipage}{0.8\linewidth}\textbf{Exercise
\arabic{exercise}:}\begin{itshape}}% begin code
{\end{itshape}\end{minipage}\end{center}}%                    end code

\newcommand{\slabel}[1]{\label{#1}}
\renewcommand{\keyword}[1]{#1}

\newcommand{\aw}{\tilde{a}} % Absorbing wall
\newcommand{\bU}{\beta}
\newcommand{\aU}{\alpha}
\newcommand{\activity}{\psi}
\newcommand{\totactivity}{\Psi}
\newcommand{\noise}{\eta}
\newcommand{\stilde}{\tilde{s}}
\newcommand{\mpos}{\tilde{\phi}} % Mapped activity, moved origin
\newcommand{\rwpos}{\phi} % Mapped activity
\newcommand{\mnoise}{\xi} % Mapped noise
\newcommand{\text}[1]{\mathrm{#1}}
\newcommand{\pdf}{\mathcal{P}}
\newcommand{\pdfS}[1]{\mathcal{P}^{\mathrm{(#1)}}}
\newcommand{\survival}[1]{\mathcal{N}^{\mathrm{(#1)}}}
\newcommand{\pdfStilde}[1]{\tilde{\mathcal{P}}^{\mathrm{(#1)}}}
\newcommand{\plaind}{\mathrm{d}}
\newcommand{\gentime}{g}
\newcommand{\KummerU}{\mathsf{U}}
\newcommand{\KummerM}{\mathsf{M}}

\newcommand{\operatorname}[1]{\,\mathrm{#1}}

\newcommand{\eqref}[1]{\eref{#1}}
\newcommand{\neref}[1]{\eref{#1}}
\newcommand{\Exp}[1]{\operatorname{exp}\left(#1\right)}
\renewcommand{\exp}[1]{\mathchoice{e^{#1}}{\operatorname{exp}\left(#1\right)}{\operatorname{exp}\left(#1\right)}{\operatorname{exp}\left(#1\right)}}
\renewcommand{\erf}[1]{\mathchoice{\mathcal{E}\left(#1\right)}{\mathcal{E}\left(#1\right)}{\mathcal{E}\left(#1\right)}{\mathcal{E}\left(#1\right)}}

\begin{document}
\title{Mapping multiplicative to additive noise}
\author{Katy J. Rubin, Gunnar Pruessner and Grigorios A. Pavliotis}
\address{Department of Mathematics,
Imperial College London,
180 Queen's Gate,
London SW7~2BZ,
UK}
\ead{gunnar.pruessner@physics.org} 
\date{Jan 2013}
\begin{abstract}
The Langevin formulation of a number of well-known stochastic processes
involves multiplicative noise. In this work we present a systematic
mapping of a process with multiplicative noise to a related process with
additive noise, which may often be easier to analyse. The mapping is
easily understood in the example of the branching process.
In a second
example we study the random neighbour (or infinite range)  contact process which is
mapped to an Ornstein-Uhlenbeck process with absorbing wall. 
The present work might
shed some light on absorbing state phase transitions in general, such as the
role of conditional expectation values and finite size scaling, 
and elucidate the
meaning of the noise amplitude.
While we
focus on the physical interpretation of the mapping, we also provide a
mathematical derivation.
\end{abstract}
\submitto{\JPA}


\maketitle

\section{Introduction}
In many stochastic processes, activity cannot recover once it has
ceased.
This feature is at the centre of absorbing state phase transitions
\cite{Hinrichsen:2000a} which in turn makes a significant part of the wider
field of non-equilibrium phase transitions
\cite{HenkelHinrichsenLuebeck:2008}.
In an equation of motion of a single degree of freedom $\activity(\gentime)$ as
a function of time $\gentime$, the feature
enters as a \keyword{multiplicative noise}, such as
\begin{equation}
\partial_{\gentime} \activity(\gentime) = f(\activity(\gentime)) +
\sqrt{\activity(\gentime)} \eta(\gentime) \ ,
\elabel{generic_eom}
\end{equation}
where 
$f(\activity)$ is a generic function with the property $f(0)=0$ and
$\eta(\gentime)$ is a noise process, to be specified below.
 Crucially, if $\activity$ vanishes at any time $\gentime^*$ it will
remain $0$ for all future times. This is exactly the feature expected
in an absorbing state phase transition and, closely related, in 
Reggeon field theory \cite{Hinrichsen:2000a,JanssenStenull:2012}.

In this paper we will consider the case where $\eta(\gentime)$ represents standard white noise and we will consider the It\={o} interpretation of the stochastic term in~\eqref{generic_eom}. This equation can be solved either numerically (or even analytically in special cases), or the corresponding
 Fokker-Planck equation that governs the evolution of the transition probability density can be studied. In the following, we will trace the origin of the multiplicative noise which
will provide an alternative formulation of the process with additive noise; as
an added benefit, non-linearities may simplify significantly, making a
direct solution of the stochastic differential equation more feasible.

It should be emphasised that a stochastic differential equation of the form~\eqref{generic_eom}, which we rewrite here for a multiplicative noise that is an arbitrary function of $\activity$
\begin{equation}
\label{e:generic_eom-1}
d \activity (\gentime) = f( \activity (\gentime)) \, dt + \sqrt{\sigma( \activity (\gentime))} \, dW (\gentime),
\end{equation} 
where $W(\gentime)$ denotes standard one-dimensional Brownian motion,
does not provide us with a complete description of the dynamics, since
noise in this equation (or, equivalently, the stochastic integral when
writing it as an integral equation) can be interpreted in different
ways, including the well known It\={o} and Stratonovich
interpretations~\cite{KSh91}. This is a modelling issue and it has to be
addressed separately. The Wong-Zakai theorem~\cite{wong} suggests that
when thinking of white noise as an idealisation of a noise process with
a non-zero correlation time, then the noise in~\eqref{e:generic_eom-1}
should be interpreted in the Stratonovich sense. On the other hand, it
is by now well known that for stochastic systems with more than one fast
time scale (one being the correlation time of the approximation to white
noise, the other being, e.g. a timescale measuring the inertia of the
system), in the limit as these timescales tending to zero, we obtain a
Stochastic Differential Equation (SDE) that can be of It\={o} or
Stratonovich type, or neither\cite{KupPavlSt04, PavSt05a}. Making the
physically correct choice of the type of noise
in~\eqref{e:generic_eom-1} is crucial, since different interpretations
of the noise lead to SDEs with qualitatively different features. A
standard example is geometric Brownian motion: the long
time behaviour of solutions to this equation (for fixed values of the
parameters in the SDE) can be different for the It\={o} and Stratonovich
SDEs~\cite{pavliotis_lecture_notes}. In this paper we will choose the
It\={o} interpretation of noise. It is well known that it is possible to
switch between different interpretations of the noise by adding an
appropriate drift (which, of course, changes generally the qualitative properties of solutions to the SDE).

Stochastic differential equations with multiplicative noise exhibit a
very rich dynamical behaviour including intermittency and noise induced
transitions~\cite{HorsLef84}. On the other hand, state-dependent noise
leads to analytical, numerical and even statistical difficulties, i.e.
it is more difficult to estimate state-dependent noise from observations
as opposed to estimating a constant diffusion coefficient; see,
e.g.~\cite{Ait-Sahalia-2008}. It is natural, therefore, to ask whether
it is possible to find an appropriate transformation that maps SDEs with
multiplicative noise to SDEs with additive noise. This is possible in
one dimension: an application of It\={o}'s formula to the function (assuming, of course, that this function exists)
\begin{equation}\label{e:lamperti}
h(x) =  \int^{x} \frac{1}{\sqrt{\sigma(\psi)}} \, d \psi,
\end{equation}
enables us to transform~\eqref{e:generic_eom-1} into an SDE with additive noise for the new process
\begin{equation}\label{e:lamperti-var}
z(\gentime) = h( \activity (\gentime)),
\end{equation}
see, e.g.~\cite{KSh91}. This transformation can also be performed at the level
of the corresponding Fokker-Planck equation, see
e.g.~\cite{MackeyLongtinLasota1990}. Such a
transformation mapping multiplicative to additive noise does not generally exist
in dimensions greater than one, unless the diffusion matrix satisfies
appropriate compatibility conditions~\cite{Ait-Sahalia-2008}.

In this paper, we adopt a different approach: we introduce a new clock,
so that, when measuring time with respect to this new time scale, noise
becomes additive. Clearly, in order for this to be possible, the
transformation to the new time must involve the actual solution of the
SDE. The theoretical basis for this random time change is the
Dambis--Dubins--Schwarz theorem~\cite[Thm 3.4.6]{KSh91}, which states
that continuous local martingales in one dimension can be expressed as
time changed Brownian motions, with the new time being the quadratic
variation of the process. For the purposes of this paper, we can state
this result as follows: we can find a new Brownian motion
$\tilde{W} (\gentime)$, such that the stochastic integral in (the integrated
version of ~\eqref{e:generic_eom-1}) can be written
\begin{equation}\label{e:dds}
\int_{0}^{t} \sqrt{\sigma(\activity (\gentime))} \, dW (\gentime) = \tilde{W} \left(\int_{0}^{t} \sigma(\activity (\gentime)) \, d \gentime \right).
\end{equation}
By construction, this change of the clock works only in one dimension.

Just as with the Lamperti transformation~\eqref{e:lamperti}, this transformation can also be performed at the level of the Fokker-Planck equation. Even though the Dambis--Dubins--Schwarz theorem is a standard result in stochastic analysis that has been used for the theoretical analysis of SDEs in one dimension (and also for the proof of homogenisation theorems with error estimates~\cite{HairPavl04}) it has not been used, to our knowledge, in the calculation of statistical quantities of interest for SDEs with multiplicative noise, in particular when boundary conditions have to be taken into account. More precisely, the connection between the change of the clock at the level of the SDE and the study of branching processes is not known. It is the goal of this paper to study precisely this problem and apply our insight to the analysis of the random neighbour contact process.

The following section motivates the mapping and 
(for illustrative purposes) exemplifies it using a continuum
formulation of the branching process.
In that case essentially all results are known in closed form. 
In \Sref{contact_process} we
proceed to apply the mapping to the random neighbour contact process, which will be turned into
an Ornstein-Uhlenbeck process with absorbing wall. \Sref{discussion} contains a
discussion of the results.

\section{Branching process}
\slabel{branching_proces}
The mapping employed in the following between a random walk (RW) and a
Watson-Galton branching process (BP),  is very well established in 
the literature
\cite{Harris:1963,Feller:1968,Feller:1966,Stapleton:2007}. We
adopt the language of a (family) tree, such as the one shown in \Fref{BP_tree}.
The branching process can be studied at
two different time scales, the slow generational time $\gentime$ (as the one
indicated on the axis in \Fref{BP_tree_tree}) and the fast individual time $t$
which corresponds to the labelling of the nodes shown in
\Fref{BP_tree_tree}. Using
the mapping as described below, the labelling within a generation and
thus the fast time scale remain somewhat arbitrary, which is irrelevant for
the argument.

\newcommand{\nope}[1]{}

\begin{figure}
\begin{subfigure}[b]{0.4\linewidth}
\begin{center}
\begin{tikzpicture}[scale=1.4]
\draw (0,0) node[circle,draw,inner sep=1pt, minimum size=20pt]  (1) {1};
\draw (1,0.8) node[circle,draw,inner sep=1pt, minimum size=20pt]  (2) {2};
\draw (1,-0.8) node[circle,draw,inner sep=1pt, minimum size=20pt]  (3) {3};
\draw (2,1.2) node[circle,draw,inner sep=1pt, minimum size=20pt] (4) {4};
\draw (2,0.4) node[circle,draw,inner sep=1pt, minimum size=20pt] (5) {5};
\draw (2,-1.2) node[circle,draw,inner sep=1pt, minimum size=20pt] (7) {6};
\draw (3,-1.5) node[circle,draw,inner sep=1pt, minimum size=20pt] (8) {\nope{11}};
\draw (3,-0.9) node[circle,draw,inner sep=1pt, minimum size=20pt] (9) {\nope{10}};
\draw (3,1.3) node[circle,draw,inner sep=1pt, minimum size=20pt] (10) {\nope{7}};
\draw (3,0.6) node[circle,draw,inner sep=1pt, minimum size=20pt] (11) {\nope{8}};
\draw (3,0.0) node[circle,draw,inner sep=1pt, minimum size=20pt] (12) {\nope{9}};
\draw (2) -- (1) -- (3);
\draw (4) -- (2) -- (5);
\draw (3) -- (7);
\draw (4) -- (10);
\draw (11) -- (5) -- (12);
\draw (9) -- (7) -- (8);
\begin{scope}[xshift=0cm,yshift=-1.9cm]
\draw [->] (-0.2,0) -- (3.5,0) node[below, at end] {$\gentime$};
\foreach \x in {0,1,2,3} {
 \draw (\x cm, 2pt) -- (\x cm, -2pt) node[below] {\small $\x$};
 }
\end{scope}
\end{tikzpicture}
\end{center}
\caption{\flabel{BP_tree_tree}Branching process as a tree.}
\end{subfigure}
\begin{subfigure}[b]{0.6\linewidth}
\begin{center}
\begin{tabular}{ccc}
$\gentime$ & $\activity(\gentime)$ & $t(\gentime)$ \\
\hline
0 & 1 & 0\\
1 & 2 & 1\\
2 & 3 & 3\\
3 & 5 & 6\\
\hspace*{0.5cm}&&
\end{tabular}
\end{center}
\caption{\flabel{BP_tree_table}Mapping of generational time $\gentime$ and individual time $t$ via the generation size $\activity(\gentime)=\rwpos(t(\gentime))$.}
\end{subfigure}
\caption{\flabel{BP_tree} Example of a branching process evolving over $3$
generations. The last generation has not yet been updated. 
The last individual which produced any offspring is the one labelled $t=6$.}
\end{figure}

A branching tree can be considered as having ``grown'' generation by
generation by
allowing each individual within a generation to reproduce. In
\Fref{BP_tree_tree} this is indicated by the labels of the nodes. If
$\activity(\gentime)$ is the number of individuals in generation
$\gentime$, the fast (microscopic) time scale $t$ may be defined as
\begin{equation}
t(\gentime) = \sum_{\gentime'=0}^{\gentime-1} \activity(\gentime) \ ,
\end{equation}
namely the total number of reproduction attempts that occurred up to 
(but excluding) generation $\gentime$, the slow (macroscopic) time scale. With obvious generalisations in
mind,
the following discussion is restricted to a
branching process with two reproduction attempts for each node, implemented
by two independent Bernoulli trials with probability $p$, \ie two offspring
are produced with probability $p^2$, a single one with $2p(1-p)$ and none
with $(1-p)^2$. In this setup, the evolution of the branching process can
be guided by a random walk of $\rwpos(t)$ which may change by
at most $1$ at two consecutive times. The population size
$\activity(\gentime+1)$ of generation
$\gentime+1$ is then determined by $\rwpos(t(\gentime+1))$ which has
taken $\activity(\gentime)$ time steps since $t(\gentime)$, so
that $\rwpos(t(\gentime))=\activity(\gentime)$. 
No further evolution can take place once $\activity(\gentime)$, or, for
that matter, $\rwpos(t)$, has vanished. In other words, the random walks
to be considered are those along an absorbing wall.

Each time step
$t$ of the random walker corresponds to a reproduction attempt of an
individual in the previous generation, as indicated by the labelling in
\Fref{BP_tree_tree}, which is not unique, yet can be interpreted as a
particular realisation of the \emph{partial} reproduction attempt of a
generation. This picture therefore affords a \keyword{bijection}
between random walk and branching process. 

\subsection{Continuum formulation}
To make further progress, the mapping is re-formulated in the continuum on
the basis of the definition
\begin{equation}
t(\gentime) = \int_0^{\gentime}\dint{\gentime'} \activity(\gentime') \ .
\elabel{time_map}
\end{equation}
and $\rwpos(t(\gentime))=\activity(\gentime)$. Since $t(0)=0$ the
initial conditions used below will be $\activity_0=\rwpos_0$.
To make \Eref{time_map} a bijection, $\activity(\gentime)$ and thus
$\rwpos(t)$ may not vanish. \Eref{time_map} is then easily inverted,
\begin{equation}
\gentime(t) = \int_0^{t}\dint{t'} \frac{1}{\rwpos(t')} \ .
\elabel{inverse_time_map}
\end{equation}

If $\rwpos(t)$ has the
equation of motion of a random walker along an absorbing wall with drift $\epsilon$,
we have
\begin{equation}
\frac{\plaind}{\plaind t} \rwpos(t) \equiv
\dot{\rwpos}(t) 
 = \epsilon + \mnoise(t)
\elabel{eom_rw}
\end{equation}
where $\mnoise(t)$ is a noise with correlator
\begin{equation}
\ave{\mnoise(t)\mnoise(t')} =
2 \Gamma^2 \delta(t-t') \ ,
\elabel{mnoise_correlator}
\end{equation}
with $\ave{\cdot}$ denoting the ensemble average.
From \Eref{eom_rw} follows
\begin{equation}
\frac{\plaind}{\plaind \gentime} \activity(\gentime) =
\frac{\plaind t}{\plaind \gentime} \dot{\rwpos}(t)
= \activity(\gentime) \left( \epsilon +  \mnoise(t(\gentime))\right)
\elabel{eom_activity_step_one}
\end{equation}
because $\frac{\plaind t}{\plaind \gentime}=\activity(\gentime)$ from
\Eref{time_map}. The equation of motion
\neref{eom_activity_step_one} for $\activity(\gentime)$
can be further simplified by introducing the noise 
\begin{equation}
\ave{\noise(\gentime)\noise(\gentime')} =
2 \Gamma^2 \delta(\gentime-\gentime')  =
2 \Gamma^2 \delta(t-t') \frac{\plaind t}{\plaind \gentime} \ ,
\elabel{noise_correlator}
\end{equation}
or, equivalently, 
\begin{equation}
\noise(\gentime) = \sqrt{\activity(\gentime)} \mnoise(t(\gentime))
\end{equation}
which results in the final continuum version of the equation of motion of
the branching process\footnote{The continuum version of the branching
process retains some
crucial features of the discrete counterpart, so that, for example,
asymptotic population sizes in the latter can be calculated on the basis of
the former using suitable effective parameters $\Gamma^2$ and $\epsilon$. This carries through even
to the total population sizes, $\totactivity$, calculated below.}
\begin{equation}
\frac{\plaind}{\plaind \gentime} \activity(\gentime) =
\activity(\gentime) \epsilon + \sqrt{\activity(\gentime)} \noise(\gentime)
\ .
\elabel{eom_bp}
\end{equation}
The term $\sqrt{\activity(\gentime)} \noise(\gentime)$ reflects the fact
that the variance of the size of each generation is linear in its previous
size (where the term ``previous'' reminds us of the It{\=o} interpretation
of \Eref{eom_bp}).

The derivation of the stochastic differential equation \neref{eom_bp} involving
multiplicative
noise from the one involving additive noise,
\Eref{eom_rw}, is invertible, \ie
\neref{time_map} implies \neref{eom_rw} given
\neref{eom_bp}. 

Mathematically, the construction above on the level of a Langevin
equation is at best handwaving: Because the
random variable $\activity(\gentime)$ enters into 
the definition \neref{time_map} of time $t$, time itself becomes a random variable. Worse, the
definition of the noise via its correlator in \Eref{noise_correlator} involves the random
variable $\plaind t/\plaind \gentime$.

At the level of a Fokker-Planck equation, the transform amounts to a
change of variables, yet unlike, say, \cite[appendix A]{BaxterBlytheMcKane:2007}, 
one of the \emph{time} variable, involving the
entire history of the random variable, \Eref{time_map}.

\subsection{Generalisation of the mapping}
\slabel{mapping_generalisation}
The mapping performed above can be generalised as follows: A Langevin
equation of the form
\begin{equation}
\frac{\plaind}{\plaind \gentime} \activity(\gentime) =
\mu(\activity(\gentime)) + \sigma(\activity(\gentime)) \noise(\gentime)
\end{equation}
with white noise $\noise(\gentime)$ as defined in \Eref{noise_correlator},
is equivalent to
\begin{equation}
\frac{\plaind}{\plaind t} \rwpos(t) =
\frac{\mu(\rwpos(t))}{\sigma^2(\rwpos(t))} + \mnoise(t)
\end{equation}
for $\rwpos(t(\gentime))=\activity(\gentime)$ along an absorbing wall,
\begin{equation}
t(\gentime) = \int_0^{\gentime} \dint{\gentime'}
\sigma^2(\activity(\gentime'))
\end{equation}
and white noise $\mnoise(t)$ as defined in \Eref{mnoise_correlator}.
However, the distribution of $\activity(\gentime)$ at fixed $\gentime$
does not equal the distribution of $\rwpos(t)$ at fixed $t$, because the
map $t(g)$ involves the history of $\activity$. Some observables,
however, do not change under the mapping and can be used to identify a
transition, which is illustrated in \Sref{comparison}.

\subsection{Fokker-Planck equations and solutions}
To understand the meaning and the consequences of the mapping introduced
above, we obtain solutions for the Fokker-Planck equations of both, the
random walk along an absorbing wall with additive noise \neref{eom_rw} and the
branching process with multiplicative noise \neref{eom_bp}.
Because for the latter the case
$\epsilon=0$ can be recovered from the general solution with
$\epsilon\ne0$ only in the form of a limit, and because $\epsilon<0$ is
qualitatively different from $\epsilon>0$, these three cases will be
discussed separately.

It is a textbook exercise to find the Fokker-Planck equation for the random
walker along an absorbing wall, \Eref{eom_rw},
which  is
\begin{equation}\elabel{FPeqn_RW}
\partial_t \pdfS{\rwpos}(\rwpos, t; \rwpos_0; \Gamma^2, \epsilon) = \Gamma^2
\partial_{\rwpos}^2 \pdfS{\rwpos} - \epsilon \partial_{\rwpos}
\pdfS{\rwpos}\ ,
\end{equation}
with $\pdfS{\rwpos}(\rwpos, t; \rwpos_0; \Gamma^2, \epsilon)$ the
probability of finding the walker at $\rwpos$ at time $t$, given it
started from $\rwpos_0$ at $t=0$,
\begin{equation}
\lim_{t\to0} \pdfS{\rwpos}(\rwpos, t; \rwpos_0; \Gamma^2,
\epsilon) = \delta(\rwpos-\rwpos_0)
\end{equation}
and given the amplitude of the noise
$\Gamma^2$, \Eref{mnoise_correlator}, and the drift $\epsilon$. 
The absorbing wall implies a Dirichlet boundary condition
\begin{equation}
\pdfS{\rwpos}(0, t; \rwpos_0; \Gamma^2, \epsilon) = 0
\end{equation}
and $\rwpos_0>0$.
The solution 
\begin{equation}
\pdfS{\rwpos}(\rwpos, t; \rwpos_0; \Gamma^2, \epsilon) =
\frac{1}{\sqrt{4 \pi \Gamma^2 t}}
\left(
\exp{-\frac{(\rwpos - \epsilon t - \rwpos_0)^2}{4 \Gamma^2 t}}
-
\exp{-\frac{(\rwpos - \epsilon t + \rwpos_0)^2}{4 \Gamma^2 t}}
\exp{-\frac{\rwpos_0 \epsilon}{\Gamma^2}}
\right)
\elabel{RW_full_PDF}
\end{equation}
is easily obtained using a mirror charge to find the solution of
\Eref{FPeqn_RW} whose $\epsilon$ is gauged away using a function $\gamma$
and writing
$\pdfS{\rwpos}=\gamma \pdfStilde{\rwpos}$, or other methods, 
\eg \cite{FarkasFulop:2001}.

The Fokker-Planck equation
of the continuous branching process, \Eref{eom_bp}, is
\begin{equation}
\partial_{\gentime} \pdfS{\activity}(\activity, \gentime; \activity_0;
\Gamma^2, \epsilon) =
\Gamma^2 \partial_{\activity}^2 \left( \activity \pdfS{\activity} \right)
- \epsilon \partial_{\activity} \left( \activity \pdfS{\activity} \right)
\elabel{full_BP_FPeqn}
\end{equation}
for $\activity>0$,
with initial condition, $\activity_0=\activity(0)=\rwpos_0$,
\begin{equation}
\lim_{\gentime\to0} \pdfS{\activity}(\activity, \gentime; \activity_0;
\Gamma^2, \epsilon) =
\delta(\activity-\activity_0) \ .
\end{equation}
Its solution is
\begin{eqnarray}
\pdfS{\activity}(\activity, \gentime; \activity_0; \Gamma^2, \epsilon) & =
&
\sqrt{\frac{\activity_0}{\activity}}
\frac{\epsilon \exp{-\epsilon \gentime/2}}%
{\Gamma^2 (1-\exp{-\epsilon \gentime})}
I_1\left(
\frac{2 \epsilon \sqrt{\activity \activity_0 \exp{-\epsilon \gentime}}}%
{\Gamma^2 (1-\exp{-\epsilon \gentime})}
\right)\nonumber\\
&&\times \Exp{-\frac{\activity \exp{-\epsilon \activity}+\activity_0}%
{\Gamma^2 (1-\exp{-\epsilon \gentime})}\epsilon} \ ,
\elabel{BP_full_PDF}
\end{eqnarray}
where $I_1$ denotes the modified Bessel function of the first kind.
The prefactor $\epsilon/(1-\exp{-\epsilon \gentime})$ has two important
properties. Firstly, for $\gentime>0$ it is positive for all
non-vanishing $\epsilon$. Secondly, taking the limit 
\begin{equation}
\lim_{\epsilon\to0} \frac{\epsilon}{1-\exp{-\epsilon \gentime}} =
\frac{1}{\gentime}
\end{equation}
recovers the solution of \Eref{full_BP_FPeqn} for $\epsilon=0$, 
\begin{equation}
\pdfS{\activity}(\activity, \gentime; \activity_0; \Gamma^2, \epsilon=0) = 
\sqrt{\frac{\activity_0}{\activity}}
\frac{1}{\Gamma^2 \gentime}
I_1\left(
\frac{2 \sqrt{\activity \activity_0}}%
{\Gamma^2 \activity}
\right)
\exp{-\frac{\activity+\activity_0}%
{\Gamma^2 \gentime}} \ .
\end{equation}
In fact, the latter solution is found in tables
\cite{Mathematica:8.0.1.0,WolframBessel-TypeFunctions:2013,GradshteynRyzhik:2000_6.615}.
From that,
\Eref{BP_full_PDF} is obtained after a sequence of transforms.
Firstly, introducing 
$\pdfS{\activity}(\activity, \gentime; \activity_0; \Gamma^2, \epsilon)
=: R(\activity \exp{-\epsilon \activity}, \activity \Gamma^2)/\activity$
simplifies \Eref{full_BP_FPeqn} to $\dot{R}(x,\stilde)=x
\exp{-\epsilon\stilde/\Gamma^2} R''(x,\stilde)$. In order to absorb the
prefactor, we introduce
$R(x,\stilde)=-(\Gamma^2/\epsilon)S(x,\exp{-\epsilon\stilde/\Gamma^2})$,
so that $\dot{S}(x,b)=-(\Gamma^2/\epsilon) x S''(x,b)$. If
$\pdfS{\activity}(\activity, \gentime; \activity_0; \Gamma^2, 0)$ solves
\Eref{full_BP_FPeqn} with $\epsilon=0$, then $S(x,b)=A x
\pdfS{\activity}(x,b-1; \activity_0, -\Gamma^2/\epsilon)$
with the same boundary and initial condition as above. The initial
condition applies at
$b=1$, the value of $\exp{-\epsilon\stilde/\Gamma^2}$ at $\stilde=0$.
Some algebra then recovers the full solution \Eref{BP_full_PDF}.

\subsection{Comparison of the branching process and the random walker
picture}
\slabel{comparison}
In the following we compare a range of observables between the different
processes or, rather, their description.
In fact, \Eref{RW_full_PDF} and \Eref{BP_full_PDF} are only two
different \emph{perspectives} on
the same process, with the advantage that one (the former, with additive
noise) is much easier
to obtain and analyse than the other. 

In the Langevin equations
\neref{eom_bp} and \neref{eom_rw}, $\activity$ and $\rwpos$
respectively do not recover from having vanished. In the following the
limits
\begin{subeqnarray}{}
\lim_{t\to\infty} \survival{\rwpos}(t;\rwpos_0; \Gamma^2, \epsilon) &=&:
\survival{\rwpos}_0(\rwpos_0; \Gamma^2, \epsilon)\\
\lim_{\gentime\to\infty} \survival{\activity}(\gentime;\activity_0; \Gamma^2, \epsilon) &=&:
\survival{\activity}_0(\activity_0; \Gamma^2, \epsilon)
\end{subeqnarray}
of the integrals
\begin{subeqnarray}{\elabel{def_survival}}
\survival{\rwpos}(t;\rwpos_0; \Gamma^2, \epsilon) &:=&
\int_0^{\infty} \dint{\rwpos} 
\pdfS{\rwpos}(\rwpos, t; \rwpos_0; \Gamma^2, \epsilon)\\ 
\survival{\activity}(\gentime;\activity_0; \Gamma^2, \epsilon) &:=&
\int_0^{\infty} \dint{\activity} 
\pdfS{\activity}(\activity, \gentime; \activity_0; \Gamma^2, \epsilon) 
\end{subeqnarray}
are referred to as the asymptotic survival probabilities. Inspecting
\Eref{time_map} shows that indefinite survival in the branching process
may not map to indefinite survival of a random walker if
$\activity(\gentime')$ vanishes fast enough. In turn, if $\rwpos(t')$
diverges quickly enough, $\gentime(t)$ in \Eref{inverse_time_map} might
remain finite in the limit $t\to\infty$. Yet, \Eref{RW_full_PDF} indicates
that the distribution of $\rwpos(t)$ is centred around
$\rwpos_0-\epsilon t$ and events beyond that scale are exponentially
suppressed. Indefinite survival of a random walker thus results in
(typically logarithmic) divergence of $\gentime(t)$ in $t$. 

We conclude
that survival of a random walker corresponds to survival of a corresponding
branching process and, likewise, early death (at finite $\gentime$) of a branching process
corresponds to an early death of a random walker (at finite $t$). One may therefore expect
that a \emph{transition} from asymptotic death to asymptotic survival in
one system corresponds to a corresponding transition in the other
system.

Integrating Equations~\eref{RW_full_PDF} and \eref{BP_full_PDF} according to \Eref{def_survival} gives
\begin{equation}\elabel{survival_RW}
\survival{\rwpos}(t;\rwpos_0; \Gamma^2, \epsilon) =
\half\left(
1-\exp{-\frac{\epsilon\rwpos_0}{\Gamma^2}}
\left(1+\erf{\frac{\epsilon t-\rwpos_0}{2\sqrt{\Gamma^2t}}} \right)
+\erf{\frac{\epsilon t+\rwpos_0}{2\sqrt{\Gamma^2t}}}
\right)
\end{equation}
and
\begin{equation}
\survival{\activity}(\gentime;\activity_0; \Gamma^2, \epsilon) =
\left\{
\begin{array}{ll}
1-\exp{-\frac{-\activity_0 \epsilon}{\Gamma^2 (1-\exp{-\epsilon \gentime})}} & \text{for}\ \ \epsilon\ne0\\
1-\exp{-\frac{-\activity_0}{\Gamma^2 \gentime}} & \text{for}\ \ \epsilon=0
\end{array}
\right.
\end{equation}
which is again continuous in $\epsilon$. Taking, however, the long time
limits gives
\begin{subeqnarray}{\elabel{survival0}}
\survival{\rwpos}_0(\rwpos_0; \Gamma^2, \epsilon) &=& 
\left\{
\begin{array}{ll}
1-\exp{-\frac{-\rwpos_0\epsilon}{\Gamma^2}} & \text{for}\  \epsilon>0\\
0 & \text{for}\  \epsilon\le0
\end{array}
\right. \\
\survival{\activity}_0(\activity_0; \Gamma^2, \epsilon) &=& 
\left\{
\begin{array}{ll}
1-\exp{-\frac{-\activity_0\epsilon}{\Gamma^2}} & \text{for}\  \epsilon>0\\
0 & \text{for}\  \epsilon\le0
\end{array}
\right. 
\end{subeqnarray}
which is,
as expected, in agreement, because $\activity_0=\rwpos_0$. A discrepancy is
however expected in the leading order behaviour
in large $t$ and $\gentime$ respectively, for
$\epsilon\le0$, which is,  
for the random walker
\begin{subnumcases}{\elabel{leading_order_survival_RW}
\survival{\rwpos}(t,\rwpos_0; \Gamma^2, \epsilon) \sim}
1-\exp{-\frac{-\rwpos_0\epsilon}{\Gamma^2}} & for  $\epsilon>0$\\
\frac{\rwpos_0}{\sqrt{\pi\Gamma^2t}}        & for  $\epsilon=0$
\elabel{leading_order_survival_RW_eps0}\\
\frac{2\rwpos_0}{(\epsilon t + \rwpos_0)(\epsilon t - \rwpos_0)}
\sqrt{\frac{\Gamma^2 t}{\pi}} \exp{-\frac{(\epsilon t + \rwpos_0)^2}{4
\Gamma^2 t}} & for $\epsilon<0$
\end{subnumcases}
and for the branching process
\begin{subnumcases}{\elabel{leading_order_survival_BP}
\survival{\activity}(\gentime,\activity_0; \Gamma^2, \epsilon) \sim}
1-\exp{-\frac{\activity_0\epsilon}{\Gamma^2}} & for  $\epsilon>0$\\
\frac{\activity_0}{\Gamma^2 \gentime} & for  $\epsilon=0$
\elabel{leading_order_survival_BP_eps0}\\
\frac{\activity_0|\epsilon|}{\Gamma^2}\exp{\epsilon \gentime} & for
$\epsilon<0$
\end{subnumcases}

The case $\epsilon=0$, obviously the critical point, deserves special
attention. It is well known 
that the survival probability in
a fair branching process, $\epsilon=0$, is inverse in the
number of generations~\cite{Harris:1963}. Since the expected
 population size remains unchanged
in the fair case, $\ave{\activity}(\gentime,\activity_0; \Gamma^2,
\epsilon=0)=\activity_0$, as discussed below, the expected population size
conditional to survival is $\Gamma^2\gentime$ according to
\Eref{leading_order_survival_BP_eps0}. On that basis,
\Eref{time_map} suggests $t(\gentime) \approx \half \Gamma^2\gentime^2$ 
which produces $\rwpos_0/(\Gamma^2\gentime\sqrt{\pi/2})$ in
\neref{leading_order_survival_RW_eps0}, out
by a factor $1/\sqrt{\pi/2}$ compared to the
asymptote for the branching process at $\epsilon=0$,
\Eref{leading_order_survival_BP_eps0}.

Other observables worth comparing are moments conditional to survival,
as they are stationary.
The moments 
\begin{equation}
\ave{\activity^n}=\int_0^\infty \dint{\activity} 
\activity^n
\pdfS{\activity}(\activity, \gentime; \activity_0; \Gamma^2, \epsilon)
\end{equation}
of \Eref{BP_full_PDF} can be calculated very easily using the
identity 
\begin{equation}
\int_0^{\infty} \dint{x} I_1(x) \exp{-\gamma x^2} = \exp{1/(4\gamma)} - 1
\end{equation}
and differentiating with respect to $\gamma$ which gives
\begin{equation}
\fl
\int_0^{\infty} \dint{\activity} \activity^{-1/2}
I_1(\sqrt{\activity}\alpha) \exp{\beta \activity} \activity^n
= 2 \alpha^{-(2n+1)}
\left.\left(
-\frac{\plaind}{\plaind \gamma}
\right)^n\right|_{\gamma=-\beta/\alpha^2}
\left(\exp{1/(4\gamma)} - 1 \right) \ .
\end{equation}
We find, for example, 
\begin{subeqnarray}{\elabel{activity_moments}}
\ave{\activity^0}&=&1-\exp{-\frac{\activity_0\epsilon}{\Gamma^2
(1-\exp{-\epsilon \gentime})}}\elabel{activity_moment0}\\
\ave{\activity^1}&=&\activity_0\exp{\epsilon\gentime}
\elabel{activity_mom1}\\
\ave{\activity^2}&=&\ave{\activity^1}^2 + 2 \activity_0
\frac{\Gamma^2}{\epsilon} \left(
\exp{2\epsilon\gentime}-\exp{\epsilon\gentime}
\right) \ .
\elabel{activity_moment2}
\end{subeqnarray}
where we use the convention
$\ave{\activity^0}=\survival{\activity}(\gentime,\activity_0; \Gamma^2,
\epsilon)$ and similarly for $\rwpos$.
Corresponding expressions for the random walker are extremely messy (as can
be seen in \Eref{survival_RW}), so we
state only the first moment (the $0$th moment is stated in
\Eref{survival_RW} and its expansion in \Eref{leading_order_survival_RW}),
dropping lower order terms in $t$ (l.o.t.):
\begin{subnumcases}{
\ave{\rwpos^1}=}
\epsilon t \left(1- 
\exp{-\frac{\epsilon\rwpos_0}{\Gamma^2}}\right) 
+ \text{l.o.t.}
& for
$\epsilon>0$\elabel{rwpos_mom1_epsgt0}\\
\rwpos_0 & for $\epsilon=0$\elabel{rwpos_mom1_eps0}\\
\frac{8\rwpos_0|\epsilon|\Gamma^2t^2}{(\epsilon t + \rwpos_0)^2(\epsilon t - \rwpos_0)^2}
\sqrt{\frac{\Gamma^2 t}{\pi}} \exp{-\frac{(\epsilon t + \rwpos_0)^2}{4
\Gamma^2 t}} 
+ \text{l.o.t.}
& for $\epsilon<0$
\end{subnumcases}
Normalising the moments with the respective survival probability
gives the moments conditional to survival, 
$\ave{\activity^n}_s=\ave{\activity^n}/\survival{\activity}$ and similar
for $\ave{\rwpos^n}_s$.
For $\epsilon>0$ the normalisation is
the same in both cases, \Eref{survival0}, so a comparison of the
conditional moments comes in fact down to comparison of the unconditional
moments. For $\activity\approx\activity_0\exp{\epsilon\gentime}$ the mapped
time is $t(\gentime)=\activity_0(\exp{\epsilon\gentime}-1)/\epsilon$ which
in \Eref{rwpos_mom1_epsgt0} gives
$\ave{\rwpos}=\activity_0(\exp{\epsilon\gentime}-1)
\left(1-\exp{-\frac{\epsilon\rwpos_0}{\Gamma^2}}\right)$,
not quite matching \Eref{activity_mom1}.

For $\epsilon=0$ the unconditional moments are identical,
\Eref{activity_mom1} (at $\epsilon=0$) and \Eref{rwpos_mom1_eps0}, so the
comparison of the conditional moments is merely a comparison of the
normalisations \neref{leading_order_survival_RW_eps0} and
\neref{leading_order_survival_BP_eps0}, which has been addressed above.

The case $\epsilon<0$ gives a conditional first moment of
$\lim_{t\to\infty}\ave{\rwpos^1}_s=\lim_{t\to\infty}\ave{\rwpos^1}/\ave{\rwpos^0}=
4\Gamma^2/|\epsilon|$ 
while the branching process gives 
$\lim_{\gentime\to\infty}\ave{\activity^1}_s=\lim_{\gentime\to\infty}
\ave{\activity^1}/\ave{\activity^0} = \Gamma^2/|\epsilon|$.

At first one may expect time independent
quantities to be 
equal in both setups.
However, as they remain subject to their dynamics, 
survivors in one system (say
the branching process) may generally be much closer to death than survivors
in the
other (say the random walkers), as they continue to linger close to extinction.

One observable that 
can be recovered 
in the random walker 
mapping in exact form is the total population size\footnote{In Self-Organised criticality
the total population size is in fact the time-integrated activity or,
equivalently, the avalanche size \cite{Pruessner:2012:Book}.} in the branching process
\begin{equation}
\totactivity = \int_0^\infty \dint{\gentime'} \activity(\gentime')
\elabel{def_totactivity}
\end{equation}
whose expectation is finite if
$\epsilon<0$ and corresponds, according to the mapping
\eref{time_map},
exactly to the time a random walker hits the absorbing wall.
This is easily confirmed for the first moment, since
\begin{equation}
\ave{\totactivity} = \int_0^\infty \dint{\gentime'}
\ave{\activity(\gentime')} = \frac{\activity_0}{|\epsilon|}
\end{equation}
for $\epsilon<0$ and because
the probability density of walkers to hit the wall at $t$ is
$(-\frac{\partial}{\partial t})\survival{\rwpos}(t;\rwpos_0; \Gamma^2,
\epsilon)$, in the random walker picture
\begin{equation}
\ave{\totactivity^n} = \int_0^\infty \dint{t}
t^n (-\frac{\partial}{\partial t}) \survival{\rwpos}(t;\rwpos_0; \Gamma^2,
\epsilon)  \ .
\elabel{totactivity_n}
\end{equation}
One thus easily finds for $\epsilon<0$
\begin{subeqnarray}{\elabel{totactivity_moments}}
\ave{\totactivity^1} & = & \frac{\rwpos_0}{|\epsilon|}
\elabel{totactivity_moment1}\\
\ave{\totactivity^2} & = & \ave{\totactivity}^2 +
\frac{2\Gamma^2\rwpos_0}{|\epsilon|^3} \\
\ave{\totactivity^3} & = & \frac{\rwpos_0}{|\epsilon^5|} 
(12\Gamma^4 + 6 \Gamma^2|\epsilon|\rwpos_0+\epsilon^2\rwpos_0^2)
\elabel{totactivity_moment3}\\
&&=
\ave{\totactivity}^3 + 3
(\ave{\totactivity^2}-\ave{\totactivity}^2)\ave{\totactivity} 
+\frac{12 \Gamma^4 \rwpos_0}{|\epsilon^5|} \nonumber
\end{subeqnarray}
obtained straight-forwardly in the random walker picture. In contrast,
deriving higher
moments of $\totactivity$ in the branching process picture is quite
cumbersome, as they require higher correlation functions, for example
\begin{equation}
\ave{\totactivity^2} = 
\int_0^\infty \dint{\gentime'}
\int_0^\infty \dint{\gentime''}
\ave{\activity(\gentime')\activity(\gentime'')} \ .
\end{equation}
In the presence of the ``kernel'' $\pdfS{\activity}$, \Eref{BP_full_PDF}, this correlation
function is determined via
\begin{equation}
\fl
\ave{\activity(\gentime')\activity(\gentime'')} = 
\int_0^\infty \dint{\activity'}
\activity' \pdfS{\activity}(\activity', \gentime'; \activity_0; \Gamma^2, \epsilon)
\int_0^\infty \dint{\activity''}
\activity'' \pdfS{\activity}(\activity'', \gentime''-\gentime'; \activity'; \Gamma^2, \epsilon)
\end{equation}
assuming $\gentime''>\gentime'$. The resulting expressions can be
evaluated using the moments calculated above,
Equations~\neref{activity_moment0}-\neref{activity_moment2}.
As expected, they exactly coincide with 
Equations~\neref{totactivity_moment1}-\neref{totactivity_moment3}.

\section{Contact process}
\slabel{contact_process}
We will now use the mapping illustrated above to characterise the
stochastic equation of motion of the random neighbour contact process.

The contact process \cite{Hinrichsen:2000a}
is a simple lattice model, for example of
the spatio-temporal evolution of an
immobile plant species spreading on a substrate. For definiteness set up
on a square lattice, sites are either occupied or empty. Occupied sites
become empty with Poissonian extinction rate $e$ and attempt to occupy with an offspring
each of their neighbouring sites with the same rate $c/q$, where
$q$ is the number of neighbours, in case of nearest neighbour
interaction, $q$ being the 
coordination number. 
Such an attempt is successful, resulting in occupation of the empty
site, 
only when the targeted site
was empty prior to the attempt. The interaction is thus due to excluded
volume, as colonisation can occur only at empty sites. Sites
become 
occupied with Poissonian colonisation rate $k c/q$, where $k$ is the
number of occupied nearest neighbours. 
Extensions to higher and lower
dimensions are obvious. Rescaling the time by $e$ determines the single
parameter controlling the dynamics as $\lambda=c/e$. It turns out
\cite{Liggett:2005} that in the thermodynamic limit a finite population density of occupied
sites $\activity$ is sustained for all $\lambda$ greater than some
$\lambda_c$, displaying all features of a second order phase transition
\cite{Fisher:1967}.

In fact, these features can already be seen in a
mean field theory, where the rise in occupation density is given by the
competition of extinction and global colonisation as a function of time
$\gentime$
\begin{equation}
\frac{\plaind}{\plaind \gentime} \activity = \lambda (1 - \activity) \activity - \activity
\elabel{CP_MFT}
\end{equation}
which has a nontrivial stationary state $\activity=1-1/\lambda>0$ for $\lambda>1$,
\ie $\lambda_c=1$. In fact, to leading order 
$\activity \propto ( \lambda - \lambda_c )^{-1}$ for $\lambda\gtrsim\lambda_c$
and $\activity=0$ otherwise.
To go beyond mean field theory, two additional
ingredients are needed, namely spatial interaction and noise.
The former is implemented by smoothing out the occupation by introducing
a diffusion term. The latter
accounts for the stochastic nature of the process.
Similar to the branching process analysed in \Sref{branching_proces},
the variance of fluctuations should be linear in the local occupation.
The full Langevin equation of
motion that is usually analysed as the contact process
\cite{Hinrichsen:2000a} reads\footnote{The absence of an explicit carrying capacity and
extinction rate spoils the usual dimensional independence of $\activity$.}
\begin{equation}
\dot{\activity}(\xvec,\gentime) = 
  \lambda (1 - \activity(\xvec,\gentime)) \activity(\xvec,\gentime) 
- \activity(\xvec,\gentime)
+ D \nabla^2 \activity
+ \sqrt{\activity(\xvec,t)}\noise(\xvec,\gentime)
\elabel{CP_full}
\end{equation}
where the noise has vanishing mean, is Gaussian, white and has correlator
\begin{equation}
\ave{\noise(\xvec,\gentime)\noise(\xvec',\gentime')} =
2 \Gamma^2 \delta(\gentime-\gentime')\delta(\xvec-\xvec') \ .
\end{equation}

This equation has been analysed extensively using field theoretic
methods \cite{Janssen:1981,Grassberger:1982,Taeuber:2005}, in
particular perturbation theory. 
Above the upper critical dimension $d>d_c=4$ 
\cite{JanssenSchaubSchmittmann:1988,BrezinZinn-Justin:1985}
spatial variation of
$\activity$ becomes irrelevant, \ie the diffusion term can be dropped,
resulting in the 
\emph{random neighbour
contact process},
\begin{equation}
\frac{\plaind}{\plaind \gentime} \activity(\gentime)= 
  \lambda (1 - \activity(\gentime)) \activity(\gentime) 
- \activity(\gentime)
+ \sqrt{\activity(\gentime)}\noise(\gentime),
\elabel{our_CP}
\end{equation}
which is the equation we will analyse in the following. 
The aim 
is to characterise \eref{our_CP}
by mapping it onto a \emph{linear} Langevin
equation with additive noise. Firstly this sheds light on the
meaning of the random neighbour model and its relation to the original,
spatial version. This will also provide a well-founded interpretation of 
the finite size scaling behaviour of the random neighbour
contact process as one would implement it numerically. Secondly, 
a number of authors have used equations of the form above to model
various natural phenomena
\cite{Levins:1969,ColizzaETAL:2006,KaluzaETAL:2010,HufnagelBrockmannGeisel:2004,Dean:1966},
such as the stochastic logistic equation, and
we expect their work to benefit directly from our analytical approach.

\subsection{Mapping the random neighbour CP to an Ornstein-Uhlenbeck process}
Using the mapping introduced in \Sref{mapping_generalisation} the original
equation of motion of the random neighbour contact process \Eref{our_CP}
can be mapped to 
\begin{equation}
\dot{\rwpos}(t) =  - \lambda  \rwpos(t) + (\lambda - 1) + \mnoise(t)  \ ,
\elabel{our_CP_mapped_V1}
\end{equation}
with an absorbing wall at $\rwpos=0$ and $t(\gentime)$ again given by \Eref{time_map}.

Was it not for the absorbing wall, the probability density
$\pdf_0(\rwpos)$ in the stationary state could be read off instantly as
the
deterministic part of \eref{our_CP_mapped_V1} can be written as
\begin{equation}
- \lambda  \rwpos(t) + (\lambda - 1) = - \frac{\dint U}{\dint \rwpos} 
\text{\ \ where\ \ }
U(\rwpos) = 
\half \lambda \left(\rwpos - \frac{\lambda-1}{\lambda} \right)^2
\elabel{potential_OU}
\end{equation}
and therefore $\pdf_0(\rwpos)\propto\exp{-U(\rwpos)/\Gamma^2}$.
The parameter $(\lambda-1)/\lambda$ translates the minimum of the
harmonic potential horizontally, while $\lambda$ itself modifies its
curvature, see \Fref{OU_potential_V1}. The dotted potential
shown there is experienced by the mirror charges placed in the system to
meet the Dirichlet boundary condition, producing a double-parabola
potential. The cusp of the potential at $\rwpos=0$ for
$\lambda\ne1$ is indicative of the technical difficulties ahead.

One may expect a phase transition due to the competition of two scales:
The distance between the absorbing wall and the minimum of the
potential, $(\lambda-1)/\lambda$, and the strength of the noise relative
to the steepness of the potential, $\Gamma/\sqrt{\lambda}$, which may or
may not drive the particles into the wall. Because the ratio of the two
lengths is dimensionless, it is possible that the transition occurs at
a non-trivial value of $\lambda$ giving rise to non-trivial exponents.

\begin{figure}
\begin{subfigure}[b]{0.33\linewidth}
\begin{tikzpicture}[scale=0.9]
\draw[very thick, domain=0:1.8, smooth, variable=\t] plot (\t,{0.5*(\t+1)*(\t+1)}) ;
\draw[very thick, dotted, domain=0:1.8, smooth, variable=\t] plot (-\t,{0.5*(\t+1)*(\t+1)}) ;
\draw[->] (-2.5,0) -- (2.5,0) node[at end, below] {$\rwpos$} ;
\draw[->] (0,-0.2) -- (0,4.2) node[at end, left] {$U$};
\draw[->] (1.50521861,2) -- (1.50521861,0) node [at end, below] {\footnotesize $\ \ave{\rwpos}_s$};
\end{tikzpicture}
\caption{$\lambda=1/2$\flabel{OU_potential_V1a}}
\end{subfigure}
\begin{subfigure}[b]{0.33\linewidth}
\begin{tikzpicture}[scale=0.9]
\draw[very thick] (0,0) parabola bend (0,0) (2,4) ;
\draw[very thick,dotted] (-2,4) parabola bend (0,0) (0,0) ;
\draw[->] (-2.5,0) -- (2.5,0) node[at end, below] {$\rwpos$} ;
\draw[->] (0,-0.2) -- (0,4.2) node[at end, left] {$U$};
\draw[->] (1.2533141373155001,2) -- (1.2533141373155001,0) node [at end, below] {\footnotesize $\ \ave{\rwpos}_s$};
\end{tikzpicture}
\caption{$\lambda=1$\flabel{OU_potential_V1b}}
\end{subfigure}
\begin{subfigure}[b]{0.33\linewidth}
\begin{tikzpicture}[scale=0.9]
\draw[very thick] (0,4/3) parabola bend (2/3,0) (1.8,3.85333) ;
\draw[very thick,dotted] (-1.8,3.8533) parabola bend (-2/3,0) (0,4/3) ;
\draw[->] (-2.5,0) -- (2.5,0) node[at end, below] {$\rwpos$} ;
\draw[->] (0,-0.2) -- (0,4.2) node[at end, left] {$U$};
\draw[->] (0.9869993558593417,2) -- (0.9869993558593417,0) node [at end, below] {\footnotesize $\ \ave{\rwpos}_s$};
\end{tikzpicture}
\caption{$\lambda=3$ \flabel{OU_potential_V1c}}
\end{subfigure}
\caption{\flabel{OU_potential_V1}The potential, \Eref{potential_OU}, for three
different values of $\lambda$. Mirror charges are subject to the dotted
potential on the left. The arrow from above marks the position of the expected
average position $\rwpos$ conditioned on survival.}
\end{figure}
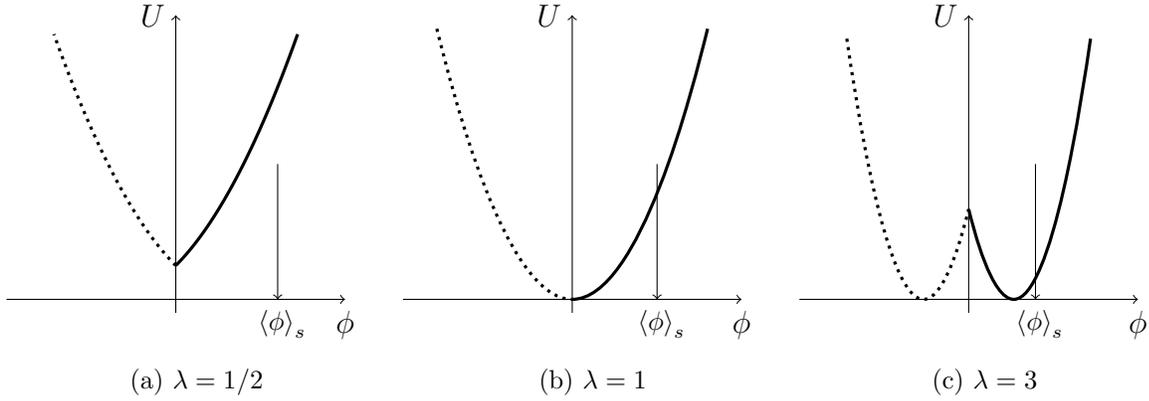

To ease notation, the origin is now moved so that the
minimum of the potential is at $\mpos=0$, resulting in the Langevin
equation analysed in the following
\begin{equation}
\dot{\mpos}(t) = - \lambda  \mpos(t) + \mnoise(t)  \ ,
\elabel{our_CP_mapped}
\end{equation}
where 
\begin{equation}
\mpos(t)=\rwpos(t)+\aw \quad \text{with} \quad
\aw=\frac{1}{\lambda}-1
\elabel{translation}
\end{equation}
the latter being the position of the absorbing wall
in a shifted potential, illustrated in
\Fref{OU_potential_V2}. The Langevin equation \neref{our_CP_mapped} is an
Ornstein-Uhlenbeck process \cite{vanKampen:1992} with an absorbing wall
\cite{Finch:2004}. 

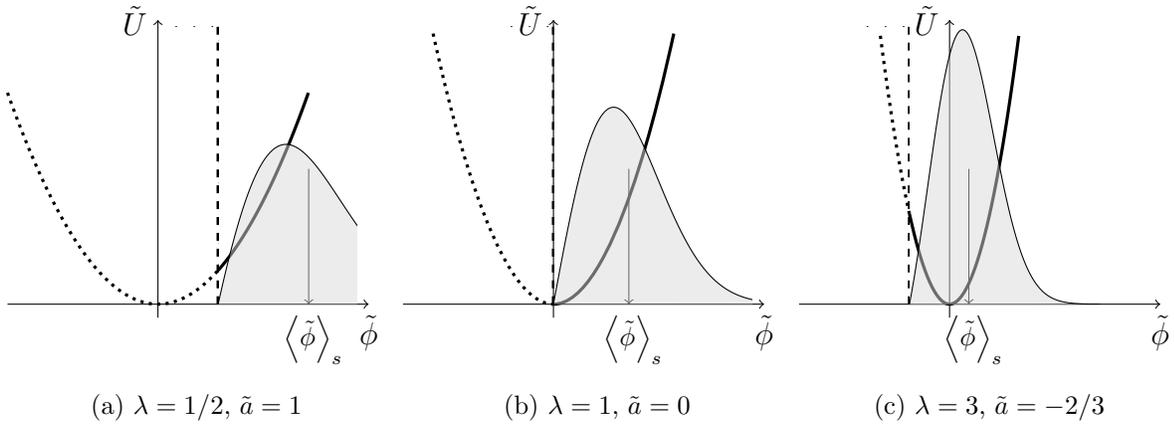
\begin{figure}[t]
\begin{subfigure}[b]{0.33\linewidth}
\begin{tikzpicture}[xscale=0.8,yscale=0.9]
\begin{scope}[shift={(1,0)}]
\clip (-1,0) rectangle (0,4.1);
\foreach \counter in {0,...,18}{ \draw (0,\counter*0.3) -- (-0.8,\counter*0.3-0.8) ; }
\path[postaction={path fading=east,fill=white,opacity=1.0}] (-0.81,0)
rectangle (0,4.1);
\draw[very thick, dashed] (0,0) -- (0,4.2) ;
\end{scope}
\draw[very thick, domain=1:2.5, smooth, variable=\t] plot
(\t,{0.5*(\t)*(\t)}) ;
\draw[very thick, dotted, domain=-2.5:1, smooth, variable=\t] plot
(\t,{0.5*(\t)*(\t)}) ;
\draw[->] (-2.5,0) -- (3.5,0) node[at end, below] {$\mpos$} ;
\draw[->] (0,-0.2) -- (0,4.2) node[at end, left] {$\tilde{U}$};
\draw[->] (1+1.50521861,2) -- (1+1.50521861,0) node [at end, below] {\footnotesize $\ \ave{\mpos}_s$};
\begin{scope}[yscale=4.8]
\clip (0,0) rectangle (3.3,4.2/4.8);
\fill[gray!30,opacity=0.5] plot file {AsymptoticPLambda_0.5.txt};
\draw plot file {AsymptoticPLambda_0.5.txt};
\end{scope}
\end{tikzpicture}
\caption{$\lambda=1/2$, $\aw=1$\flabel{OU_potential_V2a}}
\end{subfigure}
\begin{subfigure}[b]{0.33\linewidth}
\begin{tikzpicture}[xscale=0.8,yscale=0.9]
\begin{scope}[shift={(0,0)}]
\clip (-1,0) rectangle (0,4.1);
\foreach \counter in {0,...,18}{\draw (0,\counter*0.3) -- (-0.8,\counter*0.3-0.8) ; }
\path[postaction={path fading=east,fill=white,opacity=1.0}] (-0.81,0)
rectangle (0,4.1);
\draw[very thick, dashed] (0,0) -- (0,4.2) ;
\end{scope}
\draw[very thick] (0,0) parabola bend (0,0) (2,4) ;
\draw[very thick,dotted] (-2,4) parabola bend (0,0) (0,0) ;
\draw[->] (-2.5,0) -- (3.5,0) node[at end, below] {$\mpos$} ;
\draw[->] (0,-0.2) -- (0,4.2) node[at end, left] {$\tilde{U}$};
\draw[->] (1.2533141373155001,2) -- (1.2533141373155001,0) node [at end, below] {\footnotesize $\ \ave{\mpos}_s$};
\begin{scope}[yscale=4.8]
\clip (0,0) rectangle (3.3,4.2/4.8);
\fill[gray!30,opacity=0.5] plot file {AsymptoticPLambda_1.txt};
\draw plot file {AsymptoticPLambda_1.txt};
\end{scope}
\end{tikzpicture}
\caption{$\lambda=1$, $\aw=0$}
\end{subfigure}
\begin{subfigure}[b]{0.33\linewidth}
\begin{tikzpicture}[xscale=0.8,yscale=0.9]
\begin{scope}[shift={(-2/3,0)}]
\clip (-1,0) rectangle (0,4.1);
\foreach \counter in {0,...,18}{ \draw (0,\counter*0.3) -- (-0.8,\counter*0.3-0.8) ; }
\path[postaction={path fading=east,fill=white,opacity=1.0}] (-0.81,0) rectangle (0,4.1);
\draw[very thick, dashed] (0,0) -- (0,4.2) ;
\end{scope}
\draw[very thick,domain=-2/3:1.15, smooth, variable=\t] plot (\t,{3.*(\t)*(\t)}) ;
\draw[very thick,dotted, domain=-1.15:-2/3, smooth, variable=\t] plot (\t,{3.*(\t)*(\t)}) ;
\draw[->] (-2.5,0) -- (3.5,0) node[at end, below] {$\mpos$} ;
\draw[->] (0,-0.2) -- (0,4.2) node[at end, left] {$\tilde{U}$};
\draw[->] (-2/3+0.9869993558593417,2) -- (-2/3+0.9869993558593417,0) node [at end, below] {\footnotesize $\ \ave{\mpos}_s$};
\begin{scope}[yscale=4.8]
\clip (-0.66,0) rectangle (3.3,4.2/4.8);
\fill[gray!30,opacity=0.5] plot file {AsymptoticPLambda_3.txt};
\draw plot file {AsymptoticPLambda_3.txt};
\end{scope}
\end{tikzpicture}
\caption{$\lambda=3$, $\aw=-2/3$\flabel{OU_potential_V2c}}
\end{subfigure}
\caption{\flabel{OU_potential_V2}The potential 
$\tilde{U}(\mpos)=U(\rwpos)=\half \lambda \mpos^2$ (as in
\Eref{potential_OU}) 
of the Ornstein-Uhlenbeck
process \Eref{our_CP_mapped} after the shift by $\aw=-1+1/\lambda$,
\Eref{translation}. The absorbing wall is
indicated by the dashed line and the hatched region, which is not
accessible for the walker. The grey, shaded areas are the asymptotic
conditional probability densities 
\Eref{asymptotic_density_proper}. 
The arrow from above marks the position of the expected
average position $\mpos$ conditioned on survival, see
\Fref{OU_potential_V1}.}
\end{figure}

\subsection{The Fokker-Planck Equation}
In the following, we will determine the ``solution''
of \Eref{our_CP_mapped}, which is expected to be much more easily obtained
than that of the original process \eref{CP_full}.
The Fokker-Planck equation reads \cite{Zinn-Justin:1997}
\begin{equation} \elabel{fokkerplanck1}
\frac{\partial}{\partial t}\pdf(\mpos;t) =  \lambda \frac{\partial }{\partial
\mpos}\left(\mpos \pdf(\mpos;t)\right) + \Gamma^2
\frac{\partial^2}{\partial \mpos^2}\pdf(\mpos;t)
\end{equation}
with boundary condition $\pdf(\aw;t)=0$ and initial condition
$\lim_{t\to0}\pdf(\mpos;t)=\delta(\mpos-\mpos_0)$. It is obvious to attempt to write
its solution in terms of eigenfunctions $y_n(x)$ with
$x=\mpos\sqrt{\lambda}/\Gamma$ and eigenvalues $\mu_n$, say
\begin{equation}\elabel{pdf_in_efcts}
\pdf(\mpos ;t) = \sum_n e^{-\mu_n \lambda t}e^{-\frac{\lambda \mpos^2}{2 \Gamma^2}} y_n(x)
\end{equation}
where $y_n$ fulfils the eigenvalues equation
\begin{equation}\elabel{Hermite}
y_n'' - xy_n' = - \mu_n y_n \ ,
\end{equation}
with $y_n(a)=0$.
The factor $\exp {- \lambda \mpos^2/(2\Gamma^2)}$ in \eref{pdf_in_efcts}
appears quite naturally; without it, the eigenvalues $\mu_n$ were negative
and each term in the series divergent. \Eref{Hermite} is in fact the
Kolmogorov backward equation \cite{HorsLef84} of \eref{fokkerplanck1} and for
$\mu_n\in\Nset$ \eref{Hermite} is the Hermite equation. 

However, Hermite polynomials do not generally solve \eref{Hermite},
because they do not generally fulfil $y_n(\aw)=0$, except when $\aw=0$
(\ie the potential in \Fref{OU_potential_V1a} without a cusp),
where $y_n(x)=H_m(x)$ for $n=0,1,2,\ldots$ and $m=2n+1$, so that
$\mu_n=m$. This solution in odd polynomials hints at the interpretation of the
problem in terms of a mirror charge trick alluded to earlier.

The Kolmogorov backward equation \eref{Hermite} is also a
Sturm-Liouville eigenvalue problem and multiplication by the weight
function $e^{-\frac{x^2}{2}}$ converts it to standard Sturm-Liouville
form
\begin{equation}
(e^{-\frac{x^2}{2}}y_n'(x))' + \mu_n e^{-\frac{x^2}{2}}y_n(x) = 0 \ .
\end{equation}
A Sturm-Liouville eigenvalue problem has a set of eigenvalues
corresponding to a complete set of orthogonal eigenfunctions that are
square integrable with respect to the weight function
\cite{AlGwaiz:2008}. This equation is a singular Sturm-Liouville problem
because it is defined on an infinite interval and therefore an extra
condition is needed such that $\sqrt{e^{-\frac{x^2}{2}}}y_n(x)$ tends to
zero as $|x| \longrightarrow \infty$, to ensure square integrability of
$y_n(x)$. If this condition holds then a complete set of solutions can
be found. 

Solutions of \eref{Hermite} can be constructed in terms of a series by studying the
recurrence relation of its coefficients. A more efficient route is the
use of special functions, such as 
confluent hypergeometric functions also known as Kummer functions
\cite{AbramowitzStegun:1970}. These are solutions of the differential equation \cite{Slater:1960}
$x \frac{d^2y}{dx^2} + (\bU - x)\frac{dy}{dx} - \aU y = 0$
where $y(x)=\KummerM(\aU;\bU;x)$ or $y(x)=\KummerU(\aU;\bU;x)$ or any linear
combination thereof. 
A solution of the form
$y=\KummerU(-\mu_n/2;1/2;x^2/2)$ solves
equation \eref{Hermite} and satisfies the Dirichlet boundary condition at
$\aw$ for suitably chosen $\mu_n$.
It turns out that of the two independent solutions, only $\KummerU$ 
has the right asymptotic behaviour in large $x$ to guarantee square
integrability. Unfortunately, $\KummerU(-\mu_n/2;1/2;x^2/2)$ does not
solve the equation for $\aw <0$ because it has a singularity at zero
and can therefore not constitute a complete system of eigenfunctions of
the Sturm-Liouville problem.

Another differential equation to consider is the parabolic cylinder
function equation
\begin{equation} \label{eq:parabolic}
\frac{d^2y}{dx^2} + \left(\nu + \frac{1}{2} -
\frac{x^2}{4}\right)y = 0\ .
\end{equation}
The solutions of this equation \cite{Bateman:1953} are the parabolic
cylinder functions $y(x)=D_\nu(x)$, which also solve equation \eref{Hermite} when of
the form 
$y = e^{\frac{x^2}{4}}D_{\nu}(x)$, specifically
$D_{\nu}(x)=\exp{-x^2/4} H_{\nu}(x)$
for integer $\nu$. For suitable $\mu_n$ the
parabolic cylinder functions satisfy the boundary condition at $\aw$,
namely $D_{\mu_n}(\aw\sqrt{\lambda}/\Gamma)=0$,
and are analytic along the whole real line \cite{Whittaker:1902}. The
solution of \eref{fokkerplanck1} with an absorbing wall at $\aw$ is thus
\begin{equation}\elabel{solution}
\pdf(\mpos ;t) = 
e^{-\frac{\lambda (\mpos^2-\mpos_0^2)}{4 \Gamma^2}}
\frac{\sqrt{\lambda}}{\Gamma} \sum_{n=1}^\infty h_n^{-1}
e^{-\mu _n \lambda t} 
D_{\mu_n}\left(\frac{\sqrt{\lambda}}{\Gamma}\mpos\right)
D_{\mu_n}\left(\frac{\sqrt{\lambda}}{\Gamma}\mpos_0\right)
\end{equation}
where $h_n^{-1}$ is a normalisation constant,
\begin{equation}\elabel{ortho_cyl}
\int_{a\sqrt{\lambda}/\Gamma}^\infty \dint{x} D_{\mu_n}(x) D_{\mu_m}(x)
=
h_n \delta_{nm}
\end{equation}
which holds for any pair $\mu_n,\mu_m$ such that
$D_{\mu_{n,m}}(\aw\sqrt{\lambda}/\Gamma)=0$. Completeness of $D_{\mu_n}(x)$
guarantees the initial condition
$\lim_{t\to0}\pdf(\mpos;t)=\delta(\mpos-\mpos_0)$.

\subsection{Observables}
In the following the process \eref{our_CP_mapped} is characterised with
respect to the parameter $\lambda$ with the aim to identify and
characterise a phase transition. As above, the observables we are
interested in  are  the survival probability, conditional moments
and time-integrated activity. 
By construction, only the latter
can be calculated exactly in the mapping to the
Ornstein-Uhlenbeck process.

\begin{figure}
\begin{center}
\begin{tikzpicture}[xscale=3]
\foreach \x in {1,...,30}{
  \draw (\x*0.1, -0.05) -- (\x*0.1, 0.05);
}
\foreach \x in {1,...,6}{
  \pgfmathsetmacro\xtext{\x*0.5}
  \draw (\x*0.5, 0.1) -- (\x*0.5, -0.1) node [below] {$\xtext$};
}
\foreach \y in {-1,...,8}{
  \draw (-0.05/3,\y*0.5) -- (0.05/3,\y*0.5);
}
\foreach \y in {1,...,4}{
  \pgfmathsetmacro\ytext{\y}
  \draw (-0.1/3,\y) -- (0.1/3,\y) node [left=5pt] {$\ytext$};
} 
\draw[->] (-0.07,0) -- (3.1,0) node [at end,below] {$\ \ \ \lambda$};
\draw[->] (0,-0.8) -- (0,4.2);
\draw[thick, dashed, domain=0.2:3, smooth, variable=\t] plot (\t,{1/\t-1});
\draw (0.28,3.6) node {$\aw$};

\draw[very thick] plot file {mu1_times_lambda_as_a_function_of_lambda_with_small_lambda.txt};
\draw (2.3,1.3) node {$\mu_1\lambda$};

\draw[very thick,dotted] plot file {mu1_as_a_function_of_lambda.txt};
\draw (0.7,1.7) node {$\mu_1$};
\end{tikzpicture}
\end{center}
\caption{\flabel{mu1_as_a_function_of_lambda}The smallest eigenvalue
$\mu_1$ (dotted),\ie the smallest root $\mu$ of 
$D_{\mu}(\aw\sqrt{\lambda}/\Gamma)=0$, shown here for $\Gamma=1$ as a function of
$\lambda$. The asymptotic death rate of the process, $\mu_1\lambda$ (full line), does not signal a phase transition at any finite value of $\lambda$.
The position of the absorbing wall,
$\aw=1/\lambda-1$, is shown as a dashed line.
}
\end{figure}
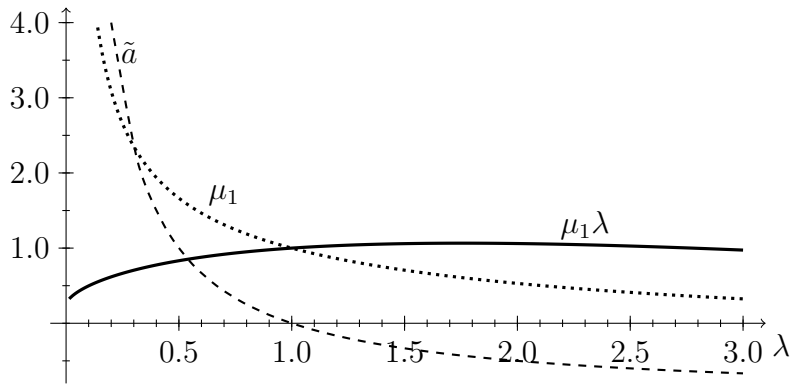

Based on the solution \eref{solution} all observables mentioned above are
easily accessible.
Asymptotically, the probability density is dominated by the smallest
eigenvalue $\mu_1$,
\begin{equation}
\elabel{asymptotic_density}
\lim_{t\to\infty} \pdf(\mpos ;t) e^{\mu_1 \lambda t}
= 
\frac{\sqrt{\lambda}}{\Gamma} h_1^{-1}
e^{-\frac{\lambda (\mpos^2-\mpos_0^2)}{4 \Gamma^2}}
D_{\mu_1}\left(\frac{\sqrt{\lambda}}{\Gamma}\mpos\right)
D_{\mu_1}\left(\frac{\sqrt{\lambda}}{\Gamma}\mpos_0\right)
\end{equation}
so that the (relative) death rate is immediately identified as $\mu_1
\lambda$, which is shown in \Fref{mu1_as_a_function_of_lambda}. 
The rate with which the system empties asymptotically, $\mu_1 \lambda$,
is positive for all positive $\lambda$. Moreover, 
the spectrum is
always discrete. In summary, (sudden) onset of asymptotic survival, as
seen in the branching process above, is not
displayed by this Ornstein-Uhlenbeck system, although this is precisely
what is expected in a contact process. This apparent clash is resolved
below.

For the following calculations a number of identities are useful, such as
the recurrence relations of the parabolic cylinder functions \cite{Bateman:1953},
\begin{equation}\elabel{cyl_ids}
\fl
\frac{\plaind}{\plaind x} \left(e^{x^2/4} D_{\nu}(x)\right) =
\nu e^{x^2/4} D_{\nu-1}(x) 
\quad \text{and}\quad 
\frac{\plaind}{\plaind x} \left(e^{-x^2/4} D_{\nu}(x)\right)  = 
- e^{-x^2/4} D_{\nu+1}(x)
\end{equation}
so that
\begin{equation}
\elabel{D_recurrence}
D_{\nu+1}(x) - x D_{\nu}(x) + \nu D_{\nu-1}(x)=0 \ .
\end{equation}

The survival probability, $\survival{\mpos}(t)= \int_{\aw}^{\infty}
\dint{\mpos}\pdf(\mpos ;t)$, therefore has asymptote
\begin{equation}\elabel{leading_order_survival_CP}
\survival{\mpos}(t,\mpos_0; \Gamma^2, \epsilon) \sim
e^{-\mu_1 \lambda t}
h_1^{-1}
e^{-\frac{\lambda (\aw^2-\mpos_0^2)}{4 \Gamma^2}}
D_{\mu_1-1}\left(\frac{\sqrt{\lambda}}{\Gamma}\aw\right)
D_{\mu_1}\left(\frac{\sqrt{\lambda}}{\Gamma}\mpos_0\right) \ .
\end{equation}
While $\survival{\mpos}$ vanishes in large $t$ for all finite $\lambda$,
the asymptotic probability density conditional to survival (see \Eref{asymptotic_density}), 
\begin{equation}\elabel{asymptotic_density_proper}
\lim_{t\to\infty}
\frac{\pdf(\mpos ;t)}{\survival{\mpos}(t)}=
\frac{\sqrt{\lambda}}{\Gamma}\exp{-\frac{\lambda(\mpos^2-\aw^2)}{4\Gamma^2}}
\frac{D_{\mu_1}(\frac{\sqrt{\lambda}\mpos}{\Gamma})}{D_{\mu_1-1}(\frac{\sqrt{\lambda}\aw}{\Gamma})}\ ,
\end{equation}
as shown in \Fref{OU_potential_V2}, is stationary, in contrast to the
branching process discussed above. As expected, the 
wall is effectively repelling, as walkers close to it are absorbed more
readily than those staying
away from it, or, conversely, if a walker survives, then because 
it stays well away from the wall.
Because $\activity$ cannot run off, but, rather, is contained
within a parabolic potential, $t(\gentime)$, \eref{time_map}, does not
necessarily diverge with $\gentime$, 
\ie
vanishing asymptotic survival probability in the random walker picture does
not imply the same in the original picture \eref{our_CP}.

With \eref{cyl_ids} and \eref{D_recurrence} 
asymptotic moments
conditional to survival are found as
\begin{equation}\elabel{ave_mpos_s}
\lim_{t\to\infty} \ave{\mpos}_s = 
\frac{\Gamma}{\sqrt{\lambda}} 
\frac{D_{\mu_1 - 2}\left(\frac{\sqrt{\lambda}}{\Gamma}\aw\right)}{D_{\mu_1 - 1}\left(\frac{\sqrt{\lambda}}{\Gamma}\aw\right)} + \aw
=
\frac{\mu_1}{\mu_1-1} \aw
\ ,
\end{equation}
and
\begin{eqnarray}\nonumber
\lim_{t\to\infty} \ave{\mpos^2}_s - \ave{\mpos}_s^2 &=&  
\frac{\Gamma^2}{\lambda} \left(
\frac{2 D_{\mu_1 - 3}\left(\frac{\sqrt{\lambda}}{\Gamma}\aw\right)}
{D_{\mu_1 - 1}\left(\frac{\sqrt{\lambda}}{\Gamma}\aw\right)}
-
\frac{D_{\mu_1 - 2}^2\left(\frac{\sqrt{\lambda}}{\Gamma}\aw\right)}
{D_{\mu_1 - 1}^2\left(\frac{\sqrt{\lambda}}{\Gamma}\aw\right)}
\right)\\
&=&\elabel{asym_var}
\frac{\aw^2\mu_1}{(\mu_1-1)^2(\mu_1-2)} - 
\frac{2 \Gamma^2}{\lambda (\mu_1-2)}
\end{eqnarray}
Because the asymptotic conditional distribution
$\pdf(\mpos ;t)/\survival{\mpos}(t)$  is stationary, one might
expect the conditional moments to be identical in the mapped and in the
original process.
To interpret the results for the contact process correctly, it is crucial to undo the shift
applied in \Eref{translation} as that affects the
mapping \Eref{inverse_time_map}. The conditional un-shifted position of
the random walker, $\lim_{t\to\infty} \ave{\rwpos}_s=\aw/(\mu_1-1)$,
must be strictly positive by construction, suggesting that
$\mu_1-1\propto \aw$ around $\lambda=1$, as $\aw$ vanishes. 

The special case $\lambda=1$ 
can be solved explicitly using Hermite polynomials, producing
\begin{equation}
\lim_{t\to\infty} \ave{\mpos}_s =
\lim_{t\to\infty} \ave{\rwpos}_s =
\sqrt{\frac{\pi}{2}\Gamma^2}
\quad \text{ for } \quad \lambda=1
\elabel{special_case}
\end{equation}
which implies $\mu_1=1-\sqrt{2/(\pi \Gamma^2)}(\lambda-1)$ to leading
order in $\lambda$ about $1$ 
and via \eref{asym_var}
\begin{equation}
\lim_{t\to\infty} \ave{\mpos^2}_s - \ave{\mpos}_s^2 =
\lim_{t\to\infty} \ave{\rwpos^2}_s - \ave{\rwpos}_s^2 =
\left(2-\frac{\pi}{2}\right) \Gamma^2
\quad \text{ for } \quad \lambda=1 \ .
\elabel{special_case_var}
\end{equation}

For very small $0<\lambda\ll1$ the potential becomes
increasingly flat while the wall $\aw$ is moving further and further to the right. 
At the wall the potential has slope $\tilde{U}'(\aw)=1-\lambda$, while
its curvature approaches
$0$. As $\mu_1$ diverges like $1/(4\Gamma^2\lambda)$
\cite{Ben-NaimKrapivsky:2010} with vanishing $\lambda$,
$\langle\mpos\rangle_s$ diverges with $\aw$ like $1/\lambda$
(the arrow in \Fref{OU_potential_V2a} moving further to the right), while
$\ave{\rwpos}_s$ converges to $4\Gamma^2$, the conditional \emph{relative}
distance to the wall as shown in \Fref{OU_potential_V1a}.

Not
unexpectedly, for very large $\lambda$ the wall has no noticeable effect,
as $\aw$ approaches $-1$ and the potential
$\tilde{U}(\mpos)=\half\lambda\mpos^2$ becomes
increasingly sharply peaked 
\begin{equation}
\lim_{t\to\infty} \ave{\mpos^2}_s - \ave{\mpos}_s^2 
\simeq
\frac{\Gamma^2}{\lambda}
\end{equation}
as if the walker were in a potential without absorbing wall, at
stationarity distributed like $\exp{-\tilde{U}(\mpos)/\Gamma^2}$.
Consequently, $\ave{\mpos}_s$
vanishes asymptotically (\Fref{OU_potential_V2c}). Given
\Eref{ave_mpos_s}, its  asymptotic
behaviour is that of $\mu_1$,
\cite{Ben-NaimKrapivsky:2010}
\begin{equation}
\mu_1 \simeq \sqrt{\frac{\lambda}{2\pi\Gamma^2}}
\exp{-\frac{\lambda}{2\Gamma^2}}
\end{equation}
Finally, the moments of the total activity in the contact process, as
defined in \Eref{totactivity_n} for the branching process,
are easily derived in the random walker picture. By the nature of the
mapping, this observable is recovered \emph{exactly}:
\begin{equation}\elabel{mom_tot_act_RNCP}
\ave{\totactivity^m} = 
e^{-\frac{\lambda (\aw^2-\mpos_0^2)}{4 \Gamma^2}}
\frac{\sqrt{\lambda}}{\Gamma} \sum_{n=1}^\infty h_n^{-1}
m! (\mu_n \lambda)^{-m}
D_{\mu_n-1}\left(\frac{\sqrt{\lambda}}{\Gamma}\aw\right)
D_{\mu_n}\left(\frac{\sqrt{\lambda}}{\Gamma}\mpos_0\right)
\end{equation}
The factor $h_n^{-1}$ on the right hand side ensures quick convergence
of the sum. There is, in fact, no suggestion that
$\ave{\totactivity^m}$ is not analytic in $\lambda$.

This concludes the calculation of the observables that are easily
derived from the random walker picture of the random neighbour contact
process. The moments of the total activity, \eref{mom_tot_act_RNCP}, are
exact, while the moments of the asymptotic conditional population
density, such as \eref{special_case} and \eref{special_case_var}, are not necessarily. None of the
observables, however, signals a transition, in contrast to the
simplified mean field theory \eref{CP_MFT}.

\section{Discussion and conclusion}
\slabel{discussion}
Before we discuss the mapping employed above in broader terms, we want
to address the question of why the random neighbour contact process, as
formulated in \Eref{our_CP}, does not display the phase transition its
mean-field theory exhibits.

The analysis above shows that in the random walker picture,
fluctuations will eventually drive the particle into the absorbing wall
irrespective of its position and the particle's starting point. Correspondingly, in the
original random neighbour contact process, the activity eventually
ceases with finite rate as long as the absorbing state is accessible. On
the other hand, every na{\"i}ve numerical implementation of the random
walker contact process will display the mean field behaviour.
Yet, numerical implementations of absorbing state phase transitions
suffer from the problem of being necessarily finite
\cite{Pruessner:2007b}. Taking the thermodynamic limit 
is crucial for the recovery of the
transition. At closer inspection, the same applies in the present model: In increasingly large systems
with volume $N$
the effective noise amplitude vanishes like $\Gamma^2\propto1/N$,
because the occupation \emph{density} $\activity$ in a large system is
less affected by the noise than in small systems.
Decreasing $\Gamma$ has the same effect on $\mu_n$ as increasing the
magnitude of
$\aw\sqrt{\lambda}$, since $D_{\mu_n}(\aw\sqrt{\lambda}/\Gamma)=0$. In
the limit of vanishing $\Gamma$ there are thus three cases: 
\begin{subnumcases}{\elabel{mu_cases}\lim_{\Gamma\to0} \mu_1 = }
\infty & for $\aw\sqrt{\lambda}>0$ \\
1      & for $\aw\sqrt{\lambda}=0$ \\
0      & for $\aw\sqrt{\lambda}<0$
\end{subnumcases}

It is obviously important to take the limit $\Gamma\to0$ in
\Eref{solution}
before considering its asymptotes in large $t$. For $\aw<0$ the particles are pinched
in an infinitely sharp potential and cannot overcome the 
barrier to the absorbing wall, \ie $\lim_{t\to\infty}\lim_{\Gamma\to0}\ave{\rwpos}=-\aw$, or
according to \Eref{ave_mpos_s}
\begin{equation}
\lim_{\Gamma\to0} \ave{\rwpos}_s = \lim_{\Gamma\to0}
\frac{1}{\mu_1-1}\aw = -\aw = 1-\frac{1}{\lambda}
\ \text{for}\ \lambda>1
\end{equation}
as $\aw\sqrt{\lambda}<0$ in \Eref{mu_cases}, reproducing the mean field result stated after \Eref{CP_MFT}.
The wall becomes accessible for $\aw\le0$ in which case
$\lim_{t\to\infty}\ave{\rwpos}_s=0$. For $\aw\sqrt{\lambda}>0$  this is in line
with \Eref{mu_cases} and \Eref{ave_mpos_s}. For $\aw=0$ the special case
\eref{special_case} applies (because $\lambda=1$) and taking the limit $\Gamma^2\to0$ there, produces again
$\ave{\rwpos}_s=0$. The limit $\Gamma\to0$ thus recovers the case
$\Gamma=0$ which leads to the mean field theory \Eref{CP_MFT} that
displays the transition.
Taking the limit 
$\Gamma\to0$ directly in \Eref{solution} using \Eref{mu_cases} is more
difficult, because we were unable to identify a suitable asymptotic
behaviour of $D_{\mu_1}(\mpos\sqrt{\lambda}/\Gamma)$ as $\mu_1$
approaches $0$ and $\Gamma$ vanishes.

In summary, the phase transition disappears provided the amplitude of
the noise correlator is finite, because all walkers
will eventually reach the absorbing wall, irrespective of the value of
$\lambda$. The transition can thus be
partly restored by studying finite $t$, 
as the characteristic time to death is given by the time scale 
$1/\lambda$
provided
by the potential,
multiplied by the distance of the wall relative to
the width of $\mpos$ produced by the noise, $\aw\sqrt{\lambda}/\Gamma$.
As $\Gamma^2\propto1/N\propto L^{-d}$ in
a $d$-dimensional system with linear extent $L$, this suggests a
dynamical exponent of $z=d/2$, which is correct at the upper critical
dimension $d=4$ \cite{Luebeck:2004}.

In the present random neighbour model the r{\^o}le of the thermodynamic
limit is thus primarily to suppress fluctuations which always drive the
system to extinction. Only once the noise has been eliminated can the
long-time limit be taken. One may expect that similar effects play a
r{\^o}le in spatially extended systems.

\subsection{Mapping multiplicative to additive noise}
The scheme introduced in \Sref{mapping_generalisation} has a very broad
range of applications. If used to cancel a multiplicative noise
amplitude, it will typically be applied with the intention to reduce the non-linearity present in the
Langevin equation. This is obviously not automatically the case. For example
\begin{equation}
\frac{\plaind}{\plaind \gentime} \activity(\gentime)= 
  \lambda (1 - \activity(\gentime)) \activity(\gentime) 
- \activity(\gentime)
+ \sqrt{\activity(\gentime)(1-\activity(\gentime))}\noise(\gentime),
\end{equation}
which is a more sophisticated version of the random neighbour contact
process that includes fluctuations in the number of unoccupied sites,
leads to the mapped equation
\begin{equation}
\dot{\rwpos}(t) =  - \lambda  + \frac{1}{1-\rwpos(t)} + \mnoise(t)  \ ,
\end{equation}
somewhat reminiscent of the Bessel process, now equipped with an absorbing wall.

In general, the scheme allows the relation of a range of different
Langevin equations and processes, some of which are much easier to
analyse than others. A potential disadvantage is that only certain
observables are exactly recovered in the mapped process; however, for a
large number of processes and mappings this is not of great significance, for
example if the primary aim is to identify a phase transition, or if the
observables are expected to be sufficiently well approximated in the mapped
process.

\section*{Acknowledgements}
The authors gratefully acknowledge very valuable discussions with
Yang Chen, 
Michael Gastner, 
Wolfram Just, 
Satya Majumdar,
Andy Parry, 
and
Hugo Touchette.

\section*{References}
\bibliography{articles,books,cp0} %,../bibtex_files/mybib}
\end{document}